# A Generalized Electrowetting Equation: Its Derivation and Consequences

Edward Bormashenko[1], Oleg Gendelman[2]

[1]*Ariel University, Physics Faculty, 40700, P.O.B. 3, Ariel, Israel*

[2]*Faculty of Mechanical Engineering, Technion – Israel Institute of Technology, Haifa 32000, Israel.*

**Abstract.**

The thermodynamics of electrowetting is treated. A general equation of electrowetting is derived from the first principles. It is demonstrated that the well-known Lippmann Equation describes a particular case of electrowetting when the radial derivative of the capacitance of the double layer is constant. The apparent contact angle of electrowetting depends on the gradient of capacity of a double layer in the vicinity of the triple line. The role of the area adjacent to the triple line in constituting the equilibrium apparent contact angle of electrowetting is emphasized.

**Introduction**

More than one century ago, Lippmann discovered that wetting can be effectively controlled by electric field [1]. Lippmann found that by applying a voltage between mercury and aqueous electrolytes, one can control the position of the mercury meniscus in a capillary [1]. The traditional scheme of electrowetting implies a direct contact of electrolyte droplet with a metal substrate, as shown in Figure 1A.

An interest in the phenomenon of electrowetting was boosted in the 1980s in a context of various applications of the effect, including lab-on-chip systems [2-3] and adaptive optical lenses [4-5]. Numerous applications of electrowetting were summarized in recent reviews [6-7]. One of the most popular modern configurations of electrowetting experiments is the so-called electrowetting-on-dielectric scheme (EWOD), depicted in Figure 1B, when liquid is placed on an insulating layer on top of bare electrodes [6-10]. At the same time the fundamentals of electrowetting have remained imperfectly understood. There exist very different and highly debatable approaches to the derivation of the Lippmann Equation of electrowetting [11-15]. The present paper derives a generalized Lippmann Equation from the first principles, and demonstrates that the apparent contact angle of electrowetting is governed by characteristics of the double layer in the nearest vicinity of the triple line.

**2. Derivation of a generalized Lippmann equation.**

In a classical electrocapillarity set-up, the phenomenon of electrowetting is related to formation of the Helmholtz double-layer at the interface between metal and electrolyte [7, 12, 16]. Charges at the interface form a parallel plate capacitor in which the gap thickness is on the order of the Debye-Huckel length. Within the modern EWOD scheme an electrolyte contacts the dielectric layer coating the metal, thus the charge separation is micrometrically scaled [6-10, 17-18]. Our treatment addresses both classical and EWOD schemes. Consider the EWOD scheme, depicted in Figures 1B and 2, when a drop partially wetting the insulating substrate is placed on the dielectric layer coating one of the electrodes. We consider electrowetting of an ideal i.e. an atomically flat, chemically homogeneous, isotropic, insoluble, non-reactive and non-deformed solid dielectric in the situation when the spreading parameter is negative [19-20]. The main macroscopic parameter describing the wetting of ideal solid substrates is the equilibrium (or Young) contact angle $\theta_Y$ [19-22]. Thus, we neglect all phenomena related to a hysteresis of the contact angle.

The free energy of a droplet $\Phi$ is supplied by the following equation [16, 19, 23]:

$$\Phi = (\gamma_{SL} - \gamma_{SA})S + \gamma S_1 + E_{vol} - \frac{C(S)\varphi^2}{2}, \qquad (1)$$

where $\gamma_{SL}$ and $\gamma_{SA}$ are the interfacial tensions at the solid-liquid and solid-air interfaces respectively, $\gamma$ is the surface tension of a liquid, $C$ is the capacitance of the double layer, $S$ is the wetted solid area before applying voltage $\varphi$, and $S_1$ is the surface of a liquid cap. $E_{vol}$ is the energy of the drop volume in an external electric field; it strongly depends on the conducting properties of the liquid, but does not depend on small variations of the contact area.

The first two terms are similar to the case of the drop without the electric field, the third term is related to the general electrical energy of the drop, and the last one is the energy reduction due to formation of the double layer. Now we write down the variation of the drop's geometric parameters in a constant external field (which is true also for a traditional electrowetting scheme, when a drop is deposited on the conducting substrate):

$$d\Phi = (\gamma_{SL} - \gamma_{SA})dS + \gamma dS_1 - \frac{\varphi^2}{2}\frac{C(S)}{dS}dS = (\gamma_{SL} + \frac{\varphi^2}{2}\frac{dC(S)}{dS} - \gamma_{SA})dS + \gamma dS_1. \quad (2)$$

Here $dS$ and $dS_1$ are the infinitesimal changes of solid-air and liquid-air surfaces, depicted in Figure 2. Let the interface tensions $\gamma_{SA}$ and $\gamma$ be independent of the external field; if we introduce then the effective surface tension $\gamma^*_{SL} = \gamma_{SL} - \frac{\varphi^2}{2}\frac{dC(S)}{dS}$ at the liquid-solid interface, the energy variation can be rewritten in the following form:

$$d\Phi = (\gamma^*_{SL} - \gamma_{SA})dS + \gamma dS_1 . \quad (3)$$

This expression is formally equivalent to the well-known variation of free energy in the absence of an electric field. Taking into account a simple geometrical relation $dS_1 = dS\cos\theta^*$ ($\theta^*$ is the equilibrium apparent contact angle of electrowetting after applying voltage, shown in Figure 2), we obtain:

$$\cos\theta^* = \frac{\gamma_{SA} - \gamma^*_{SL}}{\gamma} = \frac{\gamma_{SA} - \gamma_{SL} + \frac{\varphi^2}{2}\frac{dC(S)}{dS}}{\gamma} = \cos\theta_Y + \frac{\varphi^2}{2\gamma}\frac{dC(S)}{dS}. \quad (4)$$

Here $\theta_Y$ is the Young equilibrium angle of the liquid/solid pair, supplied by the Young Equation: $\cos\theta_Y = \frac{\gamma_{SA} - \gamma_{SL}}{\gamma}$. Eq. 4 is the equation of electrowetting. For the axisymmetrical droplet, it may be rewritten as:

$$\cos\theta^* = \cos\theta_Y + \frac{\varphi^2}{4\pi a\gamma}\frac{dC}{dr}\bigg|_{r=a}, \quad (5)$$

where *a* is the contact radius of a droplet, shown in Figure 2. Now we take a close look at Eqs. 4-5. They coincide with the well-known Lippmann Equation when the substrate is homogeneous (it may be metallic or dielectric in the EWOD scheme), i.e. $C = \tilde{C}S$ ($\tilde{C} = const$ is the specific capacitance of the unit area of a double layer). Thus, the Lippmann Equation appears in a traditional form:

$$\cos\theta^* = \cos\theta_Y + \frac{\tilde{C}\varphi^2}{2\gamma}. \qquad (6)$$

However, the electrowetting of heterogeneous surfaces such as depicted in Figure 3 is not described by Eq. 6. Consider a drop deposited on a surface manufactured from material A, comprising a central spot made from a material B, as shown in Figure 3. The equilibrium apparent contact angle of electrowetting $\theta^*$ will be governed by the surface derivative of capacitance $\frac{dC}{dr}$ taken in the vicinity of the triple line (this is typical for various wetting regimes [20, 24-27]). Material B has no influence on the equilibrium apparent contact angle $\theta^*$, as one might erroneously conclude from Eq. 6. It is noteworthy that in contrast, the energy of adhesion is influenced by the entire area wetted by a droplet [28]. Considering this reasoning is important for the design of electrowetting-driven microfluidic devices [2-3].

As it was shown recently the condition $\tilde{C} = const$ does not necessarily takes place, especially in the situation where electrowetting is accompanied by the wetting transition [29]. It is noteworthy that the triple line is usually de-pinned in the electrowetting experiments [30]; this provides an increase in the wetted area, in turn providing a non-zero surface derivative of capacitance. The increase in capacity may also arise from filling the porous substrate, as shown in Ref. 29.

The correct understanding of Eq. 4 is also important from an engineering point of view. Indeed, it is not a specific capacitance but the gradient of capacitance in the vicinity of the triple line that will govern the effect of electrowetting. Hence, for low-voltage electrowetting, high gradients of the specific capacitance appearing in the vicinity of the triple line are necessary.

In our treatment we neglected the role of the line tension [31-32] in electrowetting, discussed in Ref 33. Note that Eq. 5 resembles the well-known Neumann-Boruvka Equation (7) predicting the equilibrium contact angle of a droplet $\theta^*$ in the situation where the line tension $\Gamma$ is considered [20]:

$$\cos\theta^* = \cos\theta_Y - \frac{\Gamma}{a\gamma}, \qquad (7)$$

if $\Gamma = -\frac{\varphi^2}{4\pi}\frac{dC}{dr}\bigg|_{r=a}$ is assumed. However, this resemblance is somewhat formal, since $\Gamma$ is not a "true" line tension; in our case it depends on the applied voltage $\varphi$, whereas the true line tension is defined by the triad: solid/liquid/vapor [31-34].

The presented treatment of the problem of electrowetting was performed with a variational approach, and it is actually reduced to the principle of virtual works as demonstrated in Ref. 35. A similar result may be obtained with the more sophisticated mathematical technique of transversality conditions of the variational problem of electrowetting [15]. Both approaches reveal a special role of the area adjacent to the triple line in constituting the apparent contact angle. The presented approach is more explicit and supplies accurate expressions predicting the electrowetting equilibrium contact angle.

**Conclusions**

The well-known Lippmann Equation represents a somewhat crude approximation to the true equation of electrowetting. We demonstrate first that the equilibrium contact angle depends on the derivative of the capacitance of the double layer taken at the triple line. Only if the derivative of the capacitance of the double layer is constant, one can obtain the traditional Lippmann Equation. This correction is particularly significant, since it underlines a well-known principle of locality: the equilibrium apparent contact angle is dictated by the wetting events occurring in the nearest vicinity of the triple line, and the area far from the triple line does not influence the equilibrium contact angle. A gradient of capacitance in the vicinity of the triple line governs the equilibrium apparent contact angle of electrowetting.

**Acknowledgements**

The authors are indebted to Mrs. Ye. Bormashenko for her kind help in preparing this paper.

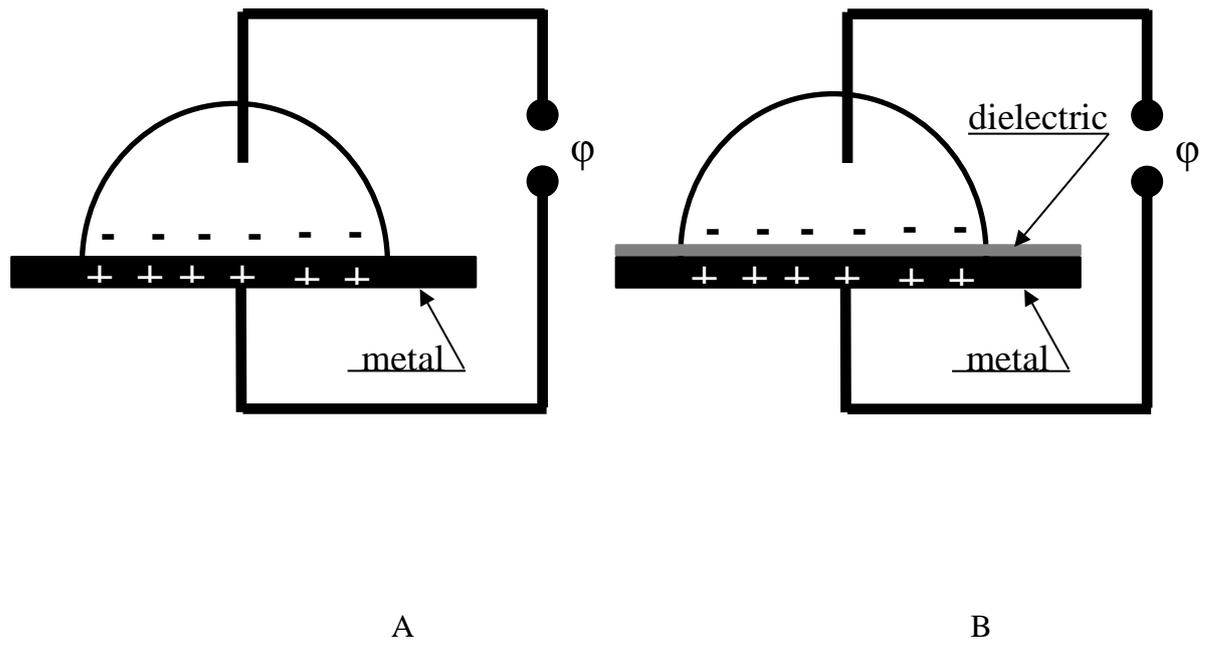

Figure 1. Traditional (A) and EWOD (B) schemes of electrowetting.

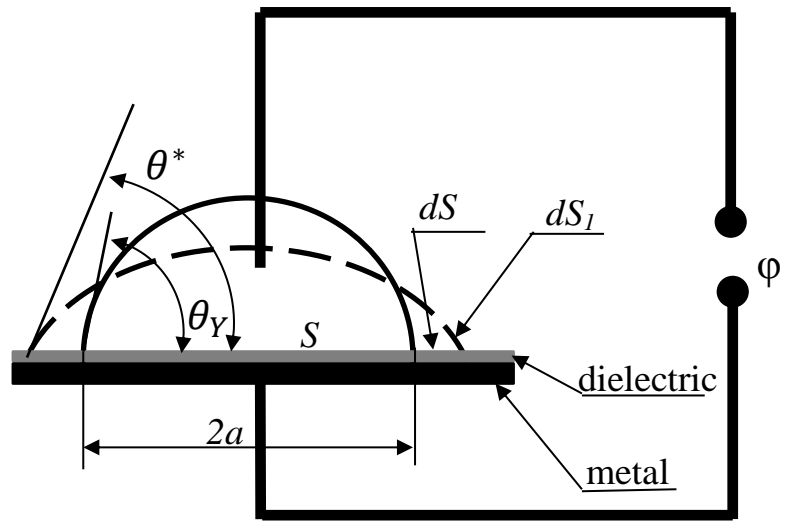

Figure 2. Parameters of electrowetting experiment under EWOD scheme.

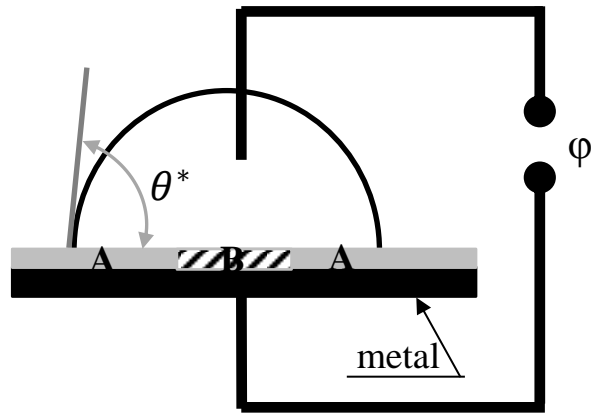

Figure 3. Electrowetting of a heterogeneous substrate built of various materials: central spot is built from material B; a droplet rests on material A.